\documentclass[twocolumn]{aa}
\usepackage[centertags]{amsmath}
\usepackage{epsfig}
\usepackage{tabularx}

\title{
	Maximal mass of uniformly rotating homogeneous stars
	in Einsteinian gravity
}
\subtitle{}
\author{K. Sch\"obel and M. Ansorg}
\offprints{K. Sch\"obel \\ \email{K.Schoebel@tpi.uni-jena.de}}
\institute{
	Theoretisch-Physikalisches Institut,
	Friedrich-Schiller-Universit\"at Jena,
	Max Wien Platz 1, 
	07743 Jena, 
	Germany
}
\date{Received / Accepted}

\abstract{
Using a multi domain spectral method, we investigate systematically the
general-relativistic model for axisymmetric uniformly rotating, homogeneous
fluid bodies generalizing the analytically known Maclaurin and Schwarzschild
solutions. Apart from the curves associated with these solutions and a further
curve of configurations that rotate at the mass shedding limit, two more
curves are found to border the corresponding two parameter set of solutions.
One of them is a Newtonian lens shaped sequence bifurcating from the Maclaurin
spheroid sequence, while the other one corresponds to highly relativistic
bodies with an infinite central pressure. The properties of the configuration
for which both the gravitational and the baryonic masses, moreover angular
velocity, angular momentum as well as polar red shift obtain their maximal
values are discussed in detail. In particular, by comparison with the static
Schwarzschild solution, we obtain an increase of 34.25\% in the gravitational
mass. Moreover, we provide exemplarily a discussion of angular velocity and
gravitational mass on the entire solution class.
}

\newcommand{\Comma}{\ensuremath{\,,}}
\newcommand{\FullStop}{\ensuremath{\,.}}

\DeclareMathOperator{\arccot}{arccot}

\begin{document}
   
\maketitle

\section{Introduction}

Various numerical methods have been developed to investigate relativistic
rotating models for extraordinarily compact astrophysical objects like neutron
stars (%
	Wilson \cite{Wilson},
	Bonazzola \& Schneider \cite{BonSchnei}, 
	Friedman et al. \cite{FriedIpsPar86, FriedIpsPar89},
	Komatsu et al. \cite{KomEriHach89a, KomEriHach89b}, 
	Lattimer et al. \cite{Latt}, 
	Neugebauer \& Herold \cite{NeuHer}, 
	Herold \& Neugebauer \cite{HerNeu}, 
	Bonazzola et al. \cite{BGSM, BonGouMarck},
	Eriguchi et al. \cite{EriHachNom94},
	Stergioulas \& Friedman \cite{SterFried},
	Nozawa et al. \cite{Nozawa}%
).
For reviews see Friedman \cite{Friedman} and Stergioulas \cite{Stergioulas}. 

The homogeneous and uniformly rotating star models were first studied by
Butterworth \& Ipser (\cite{ButIps75, ButIps76}) who found that, in addition
to the analytically known Schwarzschild and Maclaurin solutions, they are
bounded by a sequence of configurations rotating at the mass shedding limit%
\footnote{
	Due to uniform rotation, a shedding of matter sets in
	when centrifugal forces balance gravity at the equator.
	Then a cusp at the star's equatorial rim appears.
}.
The investigation of such limit configurations is instructive since certain
physical quantities reach maximal values there. Butterworth \& Ipser for
example estimated a 30\% increase in mass for fixed central pressure owing to
rotation.

In the present Letter we completed their studies, finding two further limiting
curves, in particular a sequence of stars with infinite central pressure and a
sequence of Newtonian lens-shaped configurations that bifurcates from the
Maclaurin spheroids before ending in a mass shedding limit (Bardeen
\cite{Bardeen}, Ansorg, Kleinw\"achter \& Meinel \cite{AnsKleiMei03b}). All
five limiting curves were found to circumscribe entirely the general
relativistic solution for homogeneous star models that are continuously joined
to the static Schwarzschild solution---hereafter called the `generalized
Schwarzschild solution'. This was done using the recently developed AKM method
(Ansorg, Kleinw\"achter \& Meinel \cite{AnsKleiMei02}, \cite{AnsKleiMei03c}),
which allows one to solve the Einstein equations to high accuracy even for
critical configurations. In particular we are able to determine to high
precision the extreme configuration possessing both a mass shed and infinite
central pressure. At this point several physical quantities reach their global
maxima and we provide explicit values for these.

In what follows we use units in which the speed of light and Newton's constant
of gravitation assume the value $1$. 

\section{Metric tensor and field equations}

The line element of a stationary, axisymmetric and asymptotically flat
spacetime describing a uniformly rotating perfect fluid body can be cast into
the form
\[
	ds^2=
	{\rm e}^{-2U}\left[{\rm e}^{2k}(d\rho^2+d\zeta^2)+W^2d\varphi^2\right]
		-{\rm e}^{2U}(ad\varphi+dt)^2
	\FullStop
\]
The corresponding Lewis-Papapetrou coordinates $(\rho,\zeta,\varphi,t)$ are
uniquely defined if we require continuity of the metric coefficients and their
first derivatives at the body's surface.
%
In a comoving frame, for which the metric assumes the same form with metric
functions $U'$, $k'$, $W'$ and $a'$, the relativistic Euler equation can
easily be integrated to determine the pressure $p$. For constant mass-energy
density $\mu$ this results in
\begin{equation} \label{pressure}
	p=\mu\left({\rm e}^{V_0-U'}-1\right)
	\Comma
\end{equation}
where $V_0$ is the constant surface value of $U'$.

Taking into account asymptotic flatness, boundary and transition conditions at
the surface and regularity along the rotational axis ($\rho=0$), the interior
and exterior field equations form a complete set of equations to be solved,
which is done by applying the AKM method. For a comprehensive discussion of
this multi domain spectral method see Ansorg et al. (\cite{AnsKleiMei03c}).

\section{Known static and Newtonian limits}

\subsection{Schwarzschild solution}

Solving Einstein's equation for a static homogeneous star yields the famous
Schwarzschild metric which reads in the above coordinates
\[
	{\rm e}^U=\frac{1-M/(2r)}{1+M/(2r)}
	\quad
	{\rm e}^k=1-\left(\frac M{2r}\right)^2
	\quad
	W={\rm e}^k\rho
	\quad
	a=0
\]
for the exterior ($r\equiv\sqrt{\rho^2+\zeta^2}\ge R$) and
\begin{gather*}
	{\rm e}^{U'}
	=\frac12\left[3\,
		\frac{1-M/(2R)}{1+M/(2R)}
		-\frac{1-Mr^2/(2R^3)}{1+Mr^2/(2R^3)}
	\right]
	\\
	{\rm e}^{k'}
	=\frac{\left[1+M/(2R)\right]^3}{1+Mr^2/(2R^3)}\,{\rm e}^{U'}
	\quad
	W'={\rm e}^{k'}\rho
	\quad
	a'=0
\end{gather*}
for the interior region ($0\le r\le R$). Here $M$ is the gravitational mass
and $R$ denotes the coordinate radius given implicitly through
\[
	M=\mu\frac{4\pi}3R^3\left(1+\frac M{2R}\right)^6
	\FullStop
\]

For any physically relevant solution the pressure \eqref{pressure} has to
remain finite, which is fulfilled for $R>M$. This imposes an upper bound on
the mass, namely
\[
	M<\frac4{9\sqrt{3\pi\mu}}=0.14477\ldots\frac1{\sqrt\mu} 
	\FullStop
\]

\subsection{Maclaurin spheroids}

In Newtonian gravity the problem of self gravitating rotating ideal fluids
requires solving the Poisson equation for the body's gravitational field while
satisfying the Euler-Lagrange equation governing its motion as an ideal fluid.
This leads to a free boundary value problem. A particular solution for uniform
rotation and constant mass density $\mu$ can be found by assuming the surface
to be a spheroid. One obtains the so called Maclaurin spheroids, parametrized
here by focal distance $\rho_0$ and the ratio $r_{\rm p}/r_{\rm e}$ between
polar radius $r_{\rm p}$ and equatorial radius $r_{\rm e}$.

Having computed the gravitational field, the Euler-Lagrange equation is seen to 
be satisfied for a squared angular velocity
\[
	\Omega^2=2\pi\mu\xi\left[(3\xi^2+1)\arccot\xi-3\xi\right]
	\Comma
	\quad
	\xi\equiv\left[\frac{r_{\rm e}^2}{r_{\rm p}^2}-1\right]^{-\frac12}
\]
(bottom solid curve in Fig. \ref{Omega2}). This relation holds independent of
$\rho_0$.

Note that Maclaurin spheroids exist for every $r_{\rm p}/r_{\rm e}\in[0,1]$
and $\rho_0\in[0,\infty[\,$, thus comprising a two parameter solution with
arbitrary mass for fixed $\mu$.

\subsection{First Newtonian lens sequence}

On the Maclaurin curve, an infinite series of points corresponding to
axisymmetric secular instabilities occurs, beginning at
$r_{\rm p}/r_{\rm e}=0.17126$, and accumulating in the Maclaurin disk limit
$r_{\rm p}/r_{\rm e}\to0$ (Chandrasekhar \cite{Chandrasekhar}, Bardeen
\cite{Bardeen}). They are bifurcation points of further Newtonian sequences
and correspond to singular post-Newtonian corrections (see Petroff
\cite{Petroff}).
%
The first one of these sequences is comprised of two segments that depart from
the first bifurcation point. One segment proceeds towards the Dyson rings (%
	Dyson \cite{Dyson1892, Dyson1893}, 
	Wong \cite{Wong}, 
	Kowalewsky \cite{Kowalewsky}, 
	Poincar\'e \cite{Poincare1885a, Poincare1885b, Poincare1885c},
	Eriguchi \& Sugimoto \cite{EriSug}%
) 
whereas the other one ends in a mass shedding limit (Bardeen \cite{Bardeen},
Ansorg et al. \cite{AnsKleiMei03b}). The latter we will refer to as the `first
Newtonian lens sequence' motivated by the shape of the corresponding bodies.

\section{Generalized Schwarzschild solution}

For a systematic investigation of the general case the choice of parameters is
not restricted by the numerical approach. Nevertheless it is convenient to
take non-ambiguous parameters that are restricted to a compact interval (here
$[0,1]$) in such a way that the endpoints represent limiting configurations.
Corresponding to the limits found, we selected the following magnitudes:
\begin{itemize}
%
\item Mass shed parameter 
(as defined in Ansorg et al. \cite{AnsKleiMei03c})
\[
	\beta
	\equiv-\frac{r_{\rm e}^2}{r_{\rm p}^2}\left.
		\frac{d(\zeta_{\rm s}^2)}{d(\rho^2)}
	\right|_{\rho=r_{\rm e}}
	=-\frac{r_{\rm e}}{r_{\rm p}^2}\lim_{\rho\to r_{\rm e}}
		\zeta_{\rm s}\frac{d\zeta_{\rm s}}{d\rho}
	\Comma
\]
where $\zeta_{\rm s}(\rho)$ is the function describing the surface shape.
Maclaurin and Schwarzschild bodies are characterized by $\beta=1$ and the mass
shed limit is fixed by $\beta=0$.
%
\item $\tilde p_{\rm c}\equiv p_{\rm c}/(\mu+p_{\rm c})$,
where $p_{\rm c}$ is the central pressure. Thus $\tilde p_{\rm c}=0$ stands
for the Newtonian limit where mass and hence pressure vanish and $\tilde
p_{\rm c}=1$ for the limiting configurations with infinite central pressure.
\end{itemize}
%
Additionally we will use the ratio $r_{\rm p}/r_{\rm e}$ of polar to
equatorial coordinate radius that is $1$ only for Schwarzschild
solutions.

Our numerical analysis revealed that the generalized Schwarzschild solution is
entirely bounded by the following five limiting curves (joined in the order
listed):
\begin{itemize}
\item the sequence of Schwarzschild solutions \\
($r_{\rm p}/r_{\rm e}=1$, $\beta=1$, $\tilde p_{\rm c}\in[0,1]$)
\item the Maclaurin sequence from the sphere to the first  
axisymmetric bifurcation point \\
($r_{\rm p}/r_{\rm e}\in[0.171,1]$, $\beta=1$, $\tilde p_{\rm c}=0$)
\item the first Newtonian lens sequence \\
($r_{\rm p}/r_{\rm e}\in[0.171,0.192]$, $\beta\in[0,1]$, $\tilde p_{\rm c}=0$)
\item a sequence of configurations rotating at the mass shedding limit
($r_{\rm p}/r_{\rm e}\in[0.192,0.573]$, $\beta=0$, $\tilde p_{\rm c}\in[0,1]$)
\item a sequence of configurations with infinite central pressure
($r_{\rm p}/r_{\rm e}\in[0.573,1]$, $\beta\in[0,1]$, $\tilde p_{\rm c}=1$)
\end{itemize}
%
This makes it possible to determine maximal values of all interesting physical
quantities (see section \ref{LastSection}).
%
Moreover, we can now state that on the generalized Schwarzschild solution no
quasistationary transition to a Kerr black hole is possible (in contrast to
what was found for the relativistic Dyson rings by Ansorg et al.
(\cite{AnsKleiMei03a})) and that the surface remains convex in
$\rho$-$\zeta$-coordinates.

\begin{figure}
	\epsfig{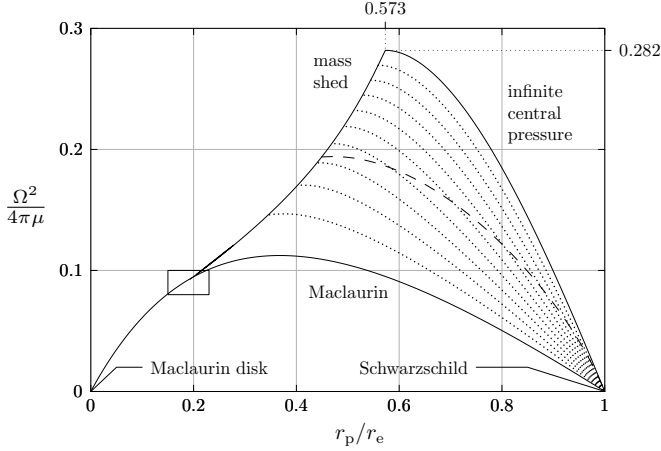}
	\caption{
		Squared angular velocity on the generalized Schwarzschild solution.
		Dotted lines indicate curves of constant central pressure 
		(equally spaced in $\tilde p_{\rm c}=p_{\rm c}/(\mu+p_{\rm c})$,
		$p_{\rm c}$ being the central pressure and
		$\mu$ the constant mass density).
		Ergo-toroids appear for configurations situated
		above the dashed line.
	}
	\label{Omega2}
\end{figure}
\begin{figure}
	\epsfig{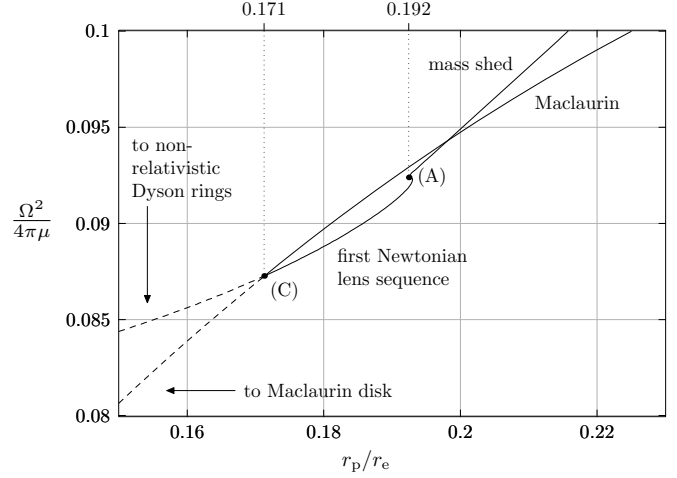}
	\caption{ 
		Magnification of Fig. \ref{Omega2}
		in the vicinity of the first axisymmetric bifurcation point (C).
		Values for the first Newtonian lens sequence were taken from 
		Ansorg et al. (\cite{AnsKleiMei03b}). 
		Points (A) and (C) correspond to the cross sections thus labelled 
		in figure 6 ibid.
	}
	\label{Zoom}
\end{figure}

Fig. \ref{Omega2} depicts the (squared) angular velocity versus the radius
ratio $r_{\rm p}/r_{\rm e}$. This is the completion of results of Butterworth
\& Ipser (\cite{ButIps75, ButIps76}).%
\footnote{
	Note that in contrast to our work they used proper radial distances
	and	kept the baryonic mass constant.
}
It shows that the configuration with maximal angular velocity rotates at the
mass shedding limit and possesses infinite central pressure. The magnification
in Fig. \ref{Zoom} demonstrates that the mass shedding curve does not
terminate at the first axisymmetric bifurcation point (C), as was conjectured
by Butterworth and Ipser. Instead it is linked to the Maclaurin spheroids at
this point via the first Newtonian lens sequence.

\begin{figure}
	\epsfig{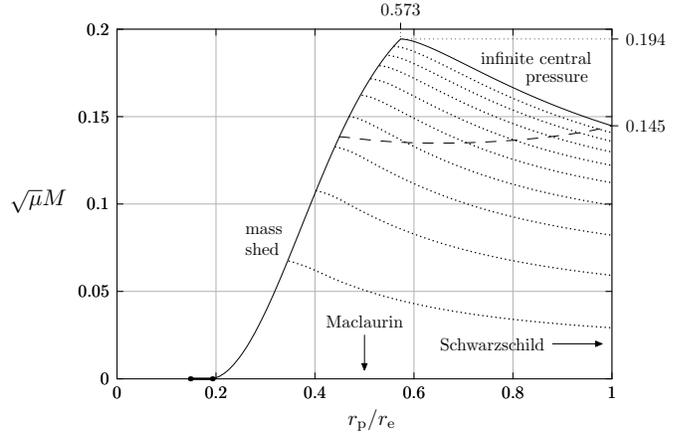}
	\caption{
		Gravitational mass on the generalized Schwarzschild solution.
		The small interval on the $r_{\rm p}/r_{\rm e}$-axis,
		emphasized by a thick line,
		represents the first Newtonian lens sequence.
		Dotted lines indicate curves of constant central pressure
		(equally spaced in $\tilde p_{\rm c}$, see Fig. \ref{Omega2}). 
		Ergo-toroids appear for configurations situated
		above the dashed line.
	}
	\label{Mass}
\end{figure}
The evolution of the gravitational mass for fixed central pressure is seen in
Fig. \ref{Mass} and reveals a 34.25\% increase in the maximal mass with
respect to the static case. The global maximum is again found for the same
configuration as for angular velocity. In both figures we included the line
above which ergo-regions appear. Interestingly, this line corresponds roughly
to one of constant mass.

Other physical quantities as baryonic mass, angular momentum, polar red shift
and circumferential radius show a similar behaviour: For fixed mass shed
parameter or radius ratio they increase with increasing central pressure.
Likewise they increase with decreasing mass shed parameter or decreasing
radius ratio if the central pressure is kept constant. So a global maximum for
each of them is obtained on the common edge of mass shed and infinite central
pressure curves, corresponding thus to a very special limit star.

\section{Maximal mass configuration}
\label{LastSection}

\begin{table}
\begin{center}
\begin{tabularx}{\linewidth}{lXr@{\,}l@{\,}lXc}
\hline
Physical quantity      &&                       & value       &              &&        \\
\hline
Gravitational mass     && $M$                   & $=0.19435 $ & $\mu^{-1/2}$ && $\ast$ \\
Baryonic mass          && $M_0$                 & $=0.27316 $ & $\mu^{-1/2}$ && $\ast$ \\
Angular velocity       && $\Omega$              & $=1.8822  $ & $\mu^{ 1/2}$ && $\ast$ \\
Angular momentum       && $J$                   & $=0.03637 $ & $\mu$        && $\ast$ \\
Polar radius           && $r_{\rm p}$           & $=0.04856 $ & $\mu^{ 1/2}$ &&        \\
Equatorial radius      && $r_{\rm e}$           & $=0.08475 $ & $\mu^{ 1/2}$ &&        \\
Radius ratio           && $r_{\rm p}/r_{\rm e}$ & $=0.5730  $ &              &&        \\
Circumferential radius && $R_\text{circ}$       & $=0.41538 $ & $\mu^{ 1/2}$ && $\ast$ \\
Polar red shift        && $Z_{\rm p}$           & $=7.378   $ &              && $\ast$ \\
\hline
\end{tabularx}
\end{center}
\caption{
	Properties of the maximal mass configuration.
	Only valid digits are given.
	An asterisk indicates that the corresponding quantity has
	a global maximum there.
}
\label{MaximalConfiguration}
\end{table}

In Table \ref{MaximalConfiguration} masses and other quantities are listed for
this extreme configuration. Because it resides on two critical curves we
notice a loss in accuracy that was somewhat recovered by an extrapolation to
an infinite approximation order $m$ of the AKM method (cf. Ansorg et al.
\cite{AnsKleiMei03c}). Observe also the unexpectedly high value for the red
shift $Z_{\rm p}$ of a photon emitted at one of the poles.

\begin{figure}
\epsfig{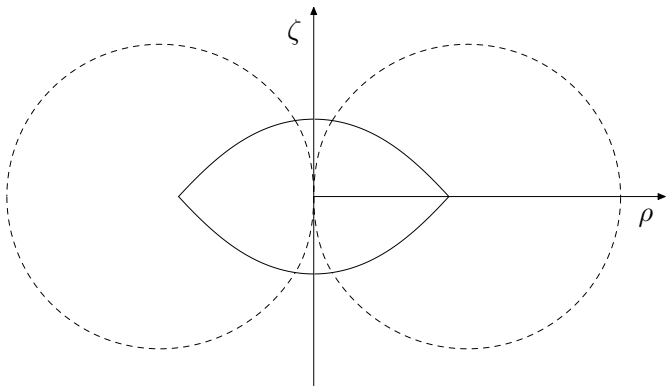}
\caption{
	Meridional coss section (solid line) and ergo-region (dashed line)
	of the maximal mass configuration. Axes are scaled identically.
}
\label{CrossSection}
\end{figure}
Fig. \ref{CrossSection} shows a meridional cross section of this configuration
including the border of the ergo-region. As in general for diverging central
pressure, the ergo-toroid degenerates by pinching together in the center.

A detailed discussion, including the relation to relativistic ring solutions,
more realistic equations of state and further going aspects like stability
will be published elsewhere.

\begin{acknowledgement}
	We would like to thank Prof. R. Meinel and D. Petroff
	for valuable discussions and helpful advice.
	This work was supported by the Deutsche Forschungsgemeinschaft
	(DFG projects ME~1820/1--3 and SFB/TR~7~--~B1).
\end{acknowledgement}

\end{document}